\documentclass{article}
\usepackage{spconf,amsmath,graphicx}
\usepackage{bm}
\usepackage{pgfplots}
\usepackage{flushend}
\usepackage[ruled,vlined]{algorithm2e}
\usepackage{tikz}
\usepackage{multirow}
\usepackage{hyperref}
\usepackage{textcomp}
\usepackage{xcolor}
\DeclareUnicodeCharacter{2212}{−}
\usepgfplotslibrary{groupplots,dateplot}
\usetikzlibrary{patterns,shapes,arrows,arrows.meta,chains,fit,positioning,calc}
\pgfplotsset{compat=newest}
\usetikzlibrary{positioning}

\newcommand{\fig}{Fig.~}
\newcommand{\tab}{Table~}

\newcommand{\desiredangle}{\alpha}
\newcommand{\edgethreshold}{\gamma}
\newcommand{\distancethreshold}{\zeta}
\newcommand{\disparitythreshold}{\beta}
\newcommand{\scanlength}{L}
\newcommand{\neighbors}{N}

\newcommand{\descoordX}{x}
\newcommand{\descoordY}{y}
\newcommand{\descoord}{\descoordX, \descoordY}
\newcommand{\filter}{f}
\newcommand{\cands}{C}
\newcommand{\sortedindices}{I}
\newcommand{\scanlines}{S}
\newcommand{\depthlinevalues}{V}
\newcommand{\warpedlines}{W}
\newcommand{\occlusionsum}{O}
\newcommand{\edgemap}{E}
\newcommand{\anglemap}{A}
\newcommand{\disparity}{D}
\newcommand{\mask}{M}

\newcommand{\T}{\text{T}}

\graphicspath{{images/}}

\AtBeginShipout{\AtBeginShipoutUpperLeft{%
		\put(\dimexpr\paperwidth-20cm\relax,-0.5cm){\parbox[t]{19cm}{\copyright 2024 IEEE. Published in 2024 International Conference on Image Processing (ICIP), scheduled for 27-30 October 2024 in Abu Dhabi, UAE. Personal use of this material is permitted. However, permission to reprint/republish this material for advertising or promotional purposes or for creating new collective works for resale or redistribution to servers or lists, or to reuse any copyrighted component of this work in other works, must be obtained from the IEEE. DOI: \url{https://doi.org/10.1109/ICIP51287.2024.10648248}}}%
}}

\title{Fast Edge-Aware Occlusion Detection in the Context of Multispectral Camera Arrays}
\name{Frank Sippel, Jürgen Seiler, and André Kaup\thanks{The authors gratefully acknowledge that this work has been supported by
the Deutsche Forschungsgemeinschaft (DFG, German Research Foundation) under project number 491814627.}}
\address{Friedrich-Alexander-Universität Erlangen-Nürnberg\\
	Multimedia Communications and Signal Processing\\
	Cauerstraße 7, 91058 Erlangen, Germany}

\begin{document}
%
\maketitle
\begin{abstract}
    Multispectral imaging is very beneficial in diverse applications, like healthcare and agriculture, since it can capture absorption bands of molecules in different spectral areas.
    A promising approach for multispectral snapshot imaging are camera arrays.
    Image processing is necessary to warp all different views to the same view to retrieve a consistent multispectral datacube.
    This process is also called multispectral image registration.
    After a cross spectral disparity estimation, an occlusion detection is required to find the pixels that were not recorded by the peripheral cameras.
    In this paper, a novel fast edge-aware occlusion detection is presented, which is shown to reduce the runtime by at least a factor of 12.
    Moreover, an evaluation on ground truth data reveals better performance in terms of precision and recall.
    Finally, the quality of a final multispectral datacube can be improved by more than 1.5 dB in terms of PSNR as well as in terms of SSIM in an existing multispectral registration pipeline.
    The source code is available at \url{https://github.com/FAU-LMS/fast-occlusion-detection}.
\end{abstract}
\section{Introduction}
\label{sec:intro}

Multispectral imaging aims at capturing more than the classical red, green and blue channel which mimic human perception.
A typical multispectral image has six to 16 different channels, often containing spectral bands in non-visible wavelength areas like infrared.
The filters of the multispectral imaging system are chosen based on the task it has to fulfill, i.e., where the molecules of interest have characteristic absorption bands.
Applications include healthcare~\cite{healthcare_2016}, agriculture~\cite{agriculture_2015}, forensics~\cite{forensics_2018} and recycling~\cite{recycling_2021}.

\begin{figure}
    \centering
    \input{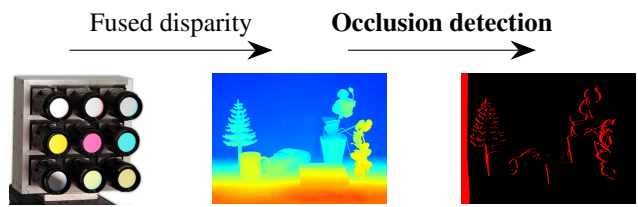}
    \caption{The problem of occlusion detection for a peripheral view from a fused disparity map tackled by this paper.}
    \label{fig:problem}
    \vspace*{-0.2cm}
\end{figure}

A promising approach for imaging a scene multispectrally, is by employing camera arrays~\cite{cam_array_1, cam_array_2, cam_array_3, cam_array_4, cam_array_5}.
There, each camera records a single spectral band.
Subsequently, image processing is required to warp all views to the same view.
This process is also called multispectral image registration.
A typical registration pipeline consists of three components~\cite{genser_camsi_2020}.
First, after calibrating the camera array, a cross spectral disparity estimator relates the same pixels of all views to each other~\cite{cade}.
Typically, the disparity map of the center camera is calculated, since it can be fused across multiple estimates, which results in a very reliable fused disparity map.
Second, after warping the peripheral views to the center view, an occlusion detection is performed to find the pixels that the center camera recorded, which are not visible in the peripheral cameras.
These two steps are summarized in \fig\ref{fig:problem}.
Third, a cross spectral reconstruction component reconstructs the occluded pixels by exploiting the structure of the center view~\cite{dgnet}.

\begin{figure*}
    \centering
    \input{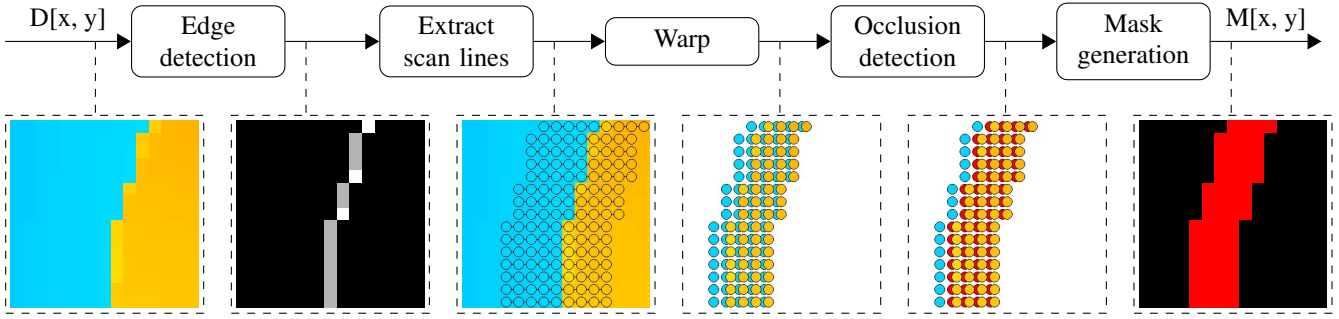}
    \caption{The proposed occlusion detection pipeline. The example images show the occlusion detection process for a horizontal stereo setup.}
    \label{fig:pipeline}
    \vspace*{-0.2cm}
\end{figure*}

In this paper, a novel fast edge-aware occlusion detection algorithm is proposed, which can easily be integrated into an existing registration pipeline.
Moreover, this occlusion detection just needs the center disparity map and thus can be used in many more camera array problems apart from multispectral camera arrays, e.g., while training disparity estimation networks~\cite{dispnet_2018}.
Note that this occlusion detection algorithm relies on disparity maps and does not work for optical flows like~\cite{of_occ_2018}, since the algorithm exploits the epipolar constraint.

\section{State of the Art}
\label{sec:sota}

The state of the art in occlusion detection is divided into neural network-based approaches and algorithmic approaches.
The neural network-based approaches usually take the left and right view of a stereo setup to estimate occlusions.
In \cite{li_symmnet_2018}, a UNet-like network~\cite{unet_2015} with skip connections is used to produce occlusion maps for left and right views.
This network architecture is a successor of the FlowNet architecture also producing occlusion maps~\cite{flownet_2015}.
An analysis of different network architectures is provided as well, which showed that this structure, where occlusions maps for both views are predicted jointly, produces the best results.
Sometimes occlusion is also estimated as byproduct of disparity estimation to improve the performance~\cite{occ_byproduct_2022}.
Neural networks have the disadvantage of reliability and interpretability.
The neural networks may miss some occluded areas, and it is hard to analyse why the neural networks failed to do so.
This behaviour makes it much more difficult to find improved architectures.

The state-of-the-art algorithmic approach by Genser et al.~\cite{genser_camsi_2020}.
This method first warps the center disparity map to the peripheral view, such that the peripheral disparity map is obtained.
During the first warping procedure, occlusions already occur, since the center disparity map does not contain all disparity values for the peripheral view.
These holes are detected by fitting alpha shapes~\cite{alpha_shapes_1994} with a radius of one pixel on the warped disparity map.
Moreover, the warped peripheral disparity map is interpolated to lie on a grid again.
This procedure is repeated into the other direction towards to center camera.
Again, holes in the warped points are detected by fitting alpha shapes~\cite{alpha_shapes_1994}.
The disadvantage of this method is that two warping steps are necessary to detect occlusions viewed from the center camera.
Moreover, it is quite slow in comparison.
In the proposed method, only one warping step is required to identify occlusions.
Moreover, this warping step is only executed on potentially occluding and occluded pixels.

\section{Edge-aware Occlusion Detection}
\label{sec:occlusion}

\begin{figure*}
    \centering
    \input{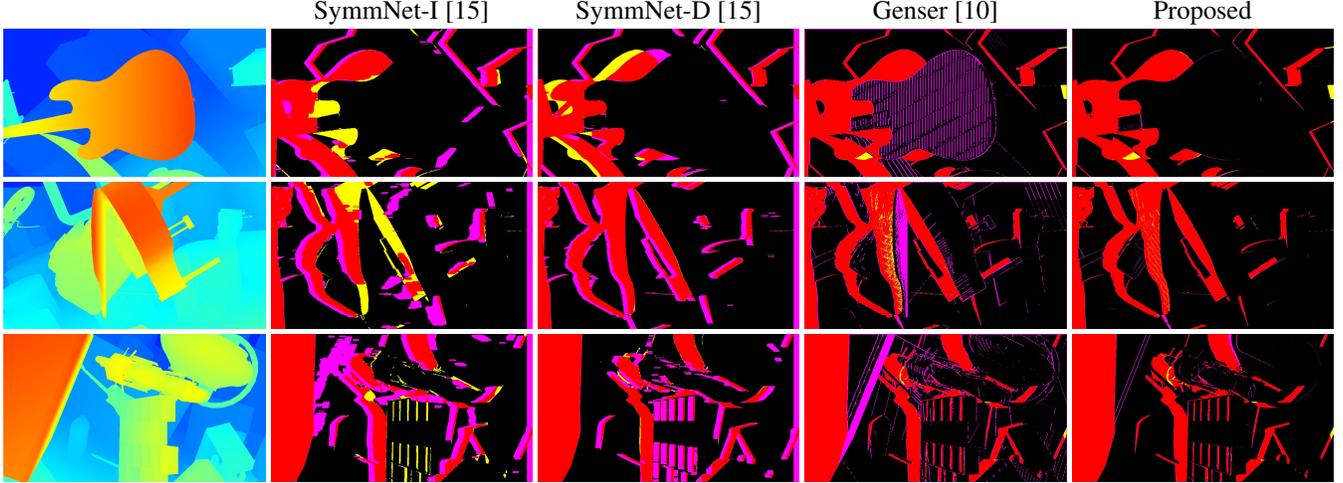}
    \caption{Examples of the result of the occlusion detection of different algorithms. A red pixel depicts a true positive (TP), a yellow pixel a false negative (FN), a purple pixel a false positive (FP) and black pixels represent true negatives (TN).}
    \label{fig:examples}
\end{figure*}

The basic idea of the proposed fast edge-aware occlusion detection is that uniform regions in the disparity map are irrelevant, since occlusions are caused by jumps in the disparity map.
For that, it is assumed that the images are calibrated and the epipolar constraint holds.
The first step is to extract edges from the disparity map by calculating the derivative using a gradient filter $\filter = (-1, 1)^\T$.
Applying this filter horizontally would be sufficient for horizontally aligned cameras.
However, since we want to apply this occlusion detection for arbitrary angles, i.e., for the array shown in \fig\ref{fig:problem}, the disparity map is also filtered vertically for these circumstances.
The horizontal and vertical edge maps are calculated by
\begin{equation}
\begin{split}
    \edgemap_x[\descoord] = \sum_i \filter[i] \cdot \disparity[x + i, y]\\
    \edgemap_y[\descoord] = \sum_i \filter[i] \cdot \disparity[x, y + i],
\end{split}
\end{equation}
respectively.
Using the horizontal and vertical edge map, the magnitude edge map $\edgemap[\descoord] = \sqrt{\edgemap_x[\descoord]^2 + \edgemap_y[\descoord]^2}$ and the angle map $\anglemap[\descoord] = \tan^{-1}\left( \frac{\edgemap_y[\descoord]}{\edgemap_x[\descoord]} \right)$ are computed.
Subsequently, these two maps are used to find the edges that cause occlusion in the direction of interest $\desiredangle$.
The list of candidates $\cands$ is found by
\begin{equation}
    \cands = \{ (x, y) \quad \text{if} \quad \edgemap[\descoord] > \edgethreshold \land |\anglemap[\descoord] - \desiredangle| < \frac{\pi}{2}\},
\end{equation}
where $\edgethreshold$ is the edge threshold representing how much the disparity map has change to cause occlusions.
The candidate list $\cands$ is then used to extract disparity map scan lines, along which occlusions are found subsequently.
These disparity map scan lines are extracted based on the angle of the baseline of the cameras $\desiredangle$.
Thus, the problem of finding occlusions effectively reduces to a one dimensional problem.
The length of these scan lines is determined by the minimum and maximum occurring disparity
\begin{equation}
    \scanlength = \lceil\left(\max(\disparity[\descoord]) - \min(\disparity[\descoord] \right) \cdot 2)\rceil.
\end{equation}
The multiplication by two is necessary since the scan lines are centered around the edges.
Starting from a general scanline {$G = [-L/2, -L/2 + 1, \dots, L/2]$}, the scan lines are calculated by considering the direction of interest $\desiredangle$
\begin{equation}
\begin{split}
    \scanlines_i^x = \cos(\desiredangle) \cdot G + \cands_i^x\\
    \scanlines_i^y = \sin(\desiredangle) \cdot G + \cands_i^y,
\end{split}
\end{equation}
where $\scanlines_i^x$ denotes the x-component of the $i$-th scanline.
Thus, the scan lines have length $\scanlength$, direction $\desiredangle$ and are centered around the corresponding candidate $\cands_i$.
Afterwards, these scan lines can be used to extract the corresponding depth values $\depthlinevalues_i = \disparity[\scanlines_i^x, \scanlines_i^y]$.
These depth values can then be used to warp the scan lines $\warpedlines_i = G_i + \depthlinevalues_i$ to the peripheral view, which is necessary to identify which pixels fall on top of each other.
To detect occlusions, these warped scan lines $\warpedlines$ are sorted, where the sorting indices are kept in $\sortedindices$, such that the corresponding depth values can still be accessed by $\depthlinevalues[\sortedindices]$.

Occlusions are detected by comparing each entry in the sorted warped lines $\warpedlines[\sortedindices]$ to its closest $\neighbors$ neighbors.
For each pair, the distance between the points and the disparity distance is checked to determine whether this pixel needs to be occluded.
For this, the number of occlusions for each pixel is calculated by
\begin{equation}
    \occlusionsum_i = \sum_{d}
    \begin{cases}
        1,      & \begin{aligned}\text{if } & |\warpedlines_i[\sortedindices] - \warpedlines_{i + d}[\sortedindices]| < \distancethreshold\ \\                               &\land \ \depthlinevalues_i[\sortedindices] < \depthlinevalues_{i + d}[\sortedindices] + \disparitythreshold
                  \end{aligned}\\
        0,      & \text{else,}
    \end{cases}
\end{equation}
where $d \in \{-\neighbors/2, ..., \neighbors/2\}/\{0\}$, $\distancethreshold$ is a distance threshold and $\disparitythreshold$ is a disparity threshold.
A smaller disparity means that this pixel belongs to the background.
Hence, if the distance between two pixels is small enough, the pixel with bigger disparity occludes the pixel with smaller disparity.
The final mask can be determined by the number of occlusions $\occlusionsum_i$ and by obeying the sorting indices
\begin{equation}
    \mask[\scanlines_i^x[I], \scanlines_i^y[I]] = \begin{cases}
        1,      & \text{if } \occlusionsum_i > 0 \\
        0,      & \text{else.}
    \end{cases}
\end{equation}
Note that the parameters $\edgethreshold, \distancethreshold$ and $\disparitythreshold$ need to determined using a training database and according to the application.
The proposed fast edge-aware occlusion detection is summarized in \fig\ref{fig:pipeline}.

\section{Evaluation}
\label{sec:evaluation}
\begin{table}
    \caption{Quantative results of different occlusion detection algorithms on the SceneFlow test set in terms of precision, recall and F score.}
    \vspace*{0.2cm}
    \label{tab:eval_occ}
    \centering
    \begin{tabular}{lcccc}
                                      & Precision & Recall & F1 Score & \\
    \hline
     SymmNet-I~\cite{li_symmnet_2018} & 0.9733 & 0.8796 & 0.9232 \\
     SymmNet-D~\cite{li_symmnet_2018} & 0.9749 & 0.9282 & 0.9496 \\
     Genser~\cite{genser_camsi_2020}  & 0.9516 & 0.9748 & 0.9628 \\
     Proposed                         & \textbf{0.9775} & \textbf{0.9781} & \textbf{0.9775}
    \end{tabular}
\end{table}

The evaluation consists of two parts.
First, the novel occlusion detection is compared to state-of-the-art methods directly on ground-truth occlusion maps.
Afterwards, the occlusion detection algorithms are evaluated using a real-world multispectral camera array, where no ground-truth occlusion maps are available, but the resulting images can be analysed.

\subsection{Occlusion Evaluation}

\begin{figure*}
    \centering
    \input{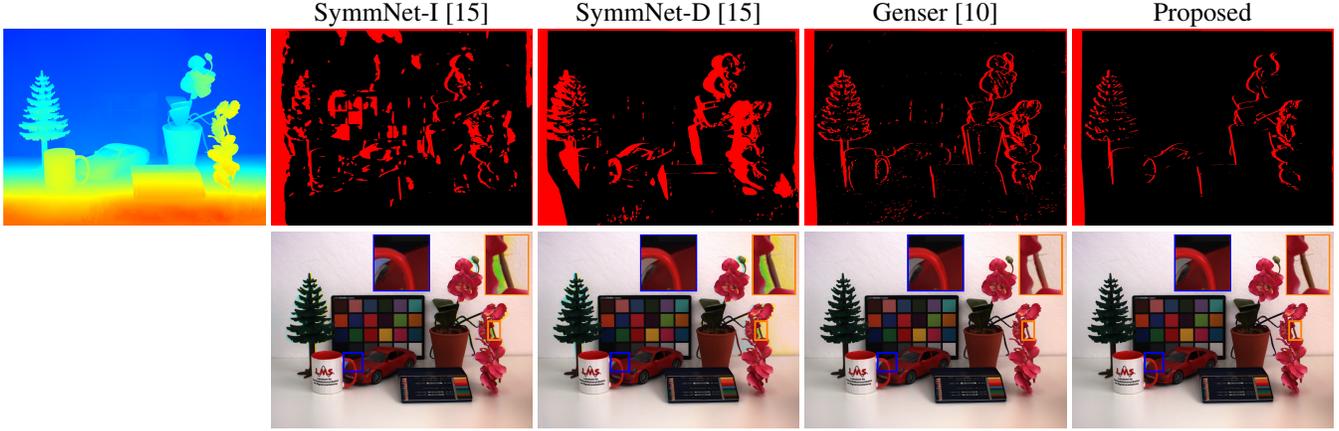}
    \caption{A real-world scene registered using the different occlusion detection algorithms.}
    \label{fig:camsi}
\end{figure*}

The first evaluation is based on the SceneFlow dataset~\cite{sceneflow_2016}.
It contains 21818 training images and 4248 validation images.
The training images were used to determine the thresholds $\edgethreshold=1.0, \distancethreshold=2.0$ and $\disparitythreshold=0.5$.
The validation images are used for the evaluation on ground-truth occlusion maps.
SymmNet~\cite{li_symmnet_2018} with stereo images as input (SymmNet-I) as well as SymmNet using the disparity map as input (SymmNet-D) were used as state-of-the-art neural network-based occlusion detection.
Furthermore, the occlusion detection by Genser et al.~\cite{genser_camsi_2020} is used as state of the art in algorithmic occluson detection.
Note that, we retrained SymmNet-D on disparity maps, since the method by Genser et al.~and our proposed method are purely based on the disparity map.
The proposed version in~\cite{li_symmnet_2018} is SymmNet-I.
The evaluation is done in terms of precision, recall and F1 score, which are calculated by
\begin{equation}
\begin{aligned}
    &\text{Precision} = \frac{\text{TP}}{\text{TP} + \text{FP}}, \quad
    \text{Recall} = \frac{\text{TP}}{\text{TP} + \text{FN}}\\
    &\text{F1 score} = \frac{2 \cdot \text{TP}}{2 \cdot \text{TP} + \text{FP} + \text{FN}}.
\end{aligned}
\end{equation}
The results for this evaluation are summarized in \tab\ref{tab:eval_occ}.
Our proposed method outperforms all other algorithm in terms of precision, recall and F1 score.
In terms of precision, our novel algorithm outperforms the method by Genser et al.~by far, while the neural network-based approaches are close.
Looking at the recall, it is vice versa.
From the occlusion detection examples in \fig\ref{fig:examples}, it gets clear that the SymmNet approaches often miss whole areas of occlusion, while the algorithmic approach by Genser et al.~tends to detect more occlusions than actually present.
Moreover, in the second example of \fig\ref{fig:examples}, one can also see that this method struggles with objects with increasing or decreasing disparity, since it has a noise-like behaviour.
The evaluations also shows that these basic neural networks work quite unreliable for this kind of task.
Furthermore, it does not get clear, in which situations the neural network produces bad predictions.
For the classic algorithms, this predictability is given much better such that it can be analysed why the method fails in which situations.
This is done in Section~\ref{sec:limitations}.

\subsection{Camera Array Evaluation}

\begin{table}
    \caption{Quantative results of different occlusion detection algorithms on a multispectral array registration task in terms of PSNR in dB and SSIM.}
    \vspace*{0.3cm}
    \label{tab:eval_camsi}
    \centering
    \begin{tabular}{@{\hspace*{0.0cm}}l@{\hspace*{0.2cm}}c@{\hspace*{0.2cm}}c@{\hspace*{0.2cm}}c@{\hspace*{0.2cm}}c@{\hspace*{0.0cm}}}
             & SymmNet-I    & SymmNet-D     & Genser        & Proposed      \\
             & \cite{li_symmnet_2018}  & \cite{li_symmnet_2018}   & \cite{genser_camsi_2020}   &               \\
    \hline
     House   & 31.43/.952 & 34.46/\textbf{.979} & 33.49/.959 & \textbf{36.12}/.977 \\
     Seaport & 42.56/.992 & 43.12/\textbf{.993} & \textbf{46.55}/\textbf{.993} & 46.50/\textbf{.993} \\
     City    & 40.29/.991 & 41.21/.993 & \textbf{46.29}/\textbf{.995} & 46.23/\textbf{.995} \\
     Outdoor & 31.14/.936 & 32.41/\textbf{.954} & 32.79/.931 & \textbf{33.82}/.944 \\
     Forest  & 25.97/.848 & 28.18/.909 & 27.15/.846 & \textbf{30.54}/\textbf{.923} \\
     Indoor  & 34.42/.982 & 34.44/.984 & 38.89/.989 & \textbf{41.76}/\textbf{.992} \\
     Lab     & 31.49/.972 & 31.58/.978 & 35.37/.979 & \textbf{36.79}/\textbf{.986} \\
    \hline
     Average & 33.90/.953 & 35.06/.970 & 37.22/.956 & \textbf{38.82}/\textbf{.973} \\
    \end{tabular}
\end{table}

The second evaluation is about registering images from the multispectral camera array shown in \fig\ref{fig:problem}.
Here, first a cross spectral disparity estimation~\cite{cade} needs to be performed to relate corresponding pixels in the different views.
This can be done for all eight peripheral cameras to the center camera as stereo disparity estimation.
In the end, these disparity maps are fused to a single disparity map by pixel-wise median filtering.
Then, after warping the image to the center view, the occlusion detection algorithms can be applied on this fused disparity map.
Finally, the occluded pixels detected are reconstructed by a cross spectral reconstruction method like~\cite{dgnet}.

This registration process can be evaluated using a synthetic hyperspectral video database~\cite{sippel_synthetic_2023}, which was rendered from a camera array very similar to the one shown in \fig\ref{fig:problem}.
This database contains seven scenes in different scenarios, each with 30 frames.
To simulate a multispectral camera array, synthetic filters have been applied.

\tab\ref{tab:eval_camsi} reveals that our proposed fast edge-aware occlusion detection also outperforms all other algorithms in this respect.
On average, our method improves the multispectral image quality by more than 1 dB PSNR in comparison to the method by Genser et al.
For a minor part of the scenes, the other methods perform slightly better either in terms of PSNR or SSIM.
To further prove this performance jump, a real-world scene with the multispectral camera has been recorded.
The results for this are shown in \fig\ref{fig:camsi}.
While the SymmNet methods contain major artifacts in the reconstructed areas, the proposed method and the method by Genser et al.~are similar.
However, when looking closer at reconstructed regions, the method by Genser et al.~suffers from some missed pixels to occlude which results in reconstruction artifacts in the final result.
These artifacts are not present in the multispectral image, where the occlusion was detected by our novel method.

\begin{table}
    \caption{Runtime of different methods in seconds.}
    \vspace*{0.3cm}
    \label{tab:runtime}
    \centering
    \begin{tabular}{lcccc}
           & SymmNet-I    & SymmNet-D     & Genser        & Proposed      \\
           & \cite{li_symmnet_2018}  & \cite{li_symmnet_2018}   & \cite{genser_camsi_2020}   &               \\
    \hline
     CPU   & 8.445 & 7.768 & 412.6 & \textbf{0.570} \\
     GPU   & 0.336 & 0.314 & X     & \textbf{0.026} \\
    \end{tabular}
\end{table}

\tab\ref{tab:runtime} shows the runtime of the different methods on CPU (Intel Core i9-7940X CPU) and GPU (NVIDIA GeForce RTX 3090).
Note that the method by Genser et al.~has no GPU implementation.
Nevertheless, the runtime can be estimated by comparing the runtime on the CPU.
The runtime measurement was executed for the real-world record by CAMSI shown in \fig\ref{fig:camsi}.
Hence, the occlusion of a 1600 \texttimes\ 1200 disparity map in eight directions was calculated by each algorithm.
The runtime evaluation shows that the proposed occlusion detection outperforms the other algorithms by far.
On the CPU, our edge-aware occlusion detection outperforms the second fastest method by a factor of over 13, while a factor of over 12 can be achieved on the GPU.
The total execution time of the novel algorithm is 26 ms for all eight images, thus making it suitable for real-time applications.

\section{Limitations}
\label{sec:limitations}

As shown, the proposed occlusion detection works reliable in most situations.
However, this type of occlusion detection has some minor limitations.
First, if an object is very close to one camera such that it is not visible in the other camera, the disparity estimation and thus also the occlusion detection will not be able to detect that.
Second, objects right outside the border of the center image will cause occlusions in the peripheral view.
These occlusions are not detectable using the center disparity map.
This can be observed in the first and third example in \fig\ref{fig:examples}.
Finally, occlusions caused by occluded objects in the center view cannot be reliably detected by any approach.
Actually, this effect is visible in the first example in \fig\ref{fig:examples} as well.
The guitar occludes the object behind it which then occludes the background.
However, since the disparity values of the second object are not provided to the occlusion detection algorithm, this occlusion cannot be detected and thus results in not properly found pixels.
This cannot be reliably identified by any algorithm, since the disparity values needed for this are not provided.
However, a neural network could infer the shape of the occluded object, and thus hallucinate the occluded occlusion.

\section{Summary}
\label{sec:summary}

In this paper, a fast edge-aware occlusion detection was presented in the context of a multispectral camera array.
For such arrays, several disparity maps are estimated beforehand to yield a fused disparity map for the center camera.
Then, this fused disparity map is used to detect occlusions in the peripheral views to determine which pixels are not visible and thus need to reconstructed.
The proposed algorithm is based on the idea that uniform regions in the disparity map are not interesting for the occlusion detection, which drastically speeds up computation time.
The occlusion detection itself is based on the warped mesh for each resulting edge pixel to look at.
The evaluation revealed that it performs better on ground-truth occlusion data in comparison to the state of the art.
Moreover, the occlusion detection was also embedded in a multispectral registration process yielding the best SSIM and a more than 1.5 dB higher PSNR in comparison to its best competitor.
Finally, also the real-time capability of the novel occlusion detection was proven.
Note that this occlusion detection algorithm can be applied to other tasks than multispectral camera arrays, as well.
In the future, further runtime reduction might be achieved by thinning down the candidate list and the number of comparisons.
Moreover, there is a need for more reliable neural network-based approaches.


\bibliographystyle{IEEEbib}
\bibliography{refs}

\end{document}